\documentclass[pre,showpacs,twocolumn]{revtex4}
\usepackage{dcolumn}
\usepackage{bm}
\usepackage{mathrsfs}
\usepackage{amssymb}
\usepackage{amsmath}
\usepackage{amsfonts}
\usepackage{color}
\usepackage{ulem}

\usepackage[]{graphicx}
\begin{document}
\newcommand{\half}{\frac{1}{2}}
\title{A mechanical analog of quantum bradyons and tachyons }
\author{Aur\'elien Drezet$^{1}$, Pierre Jamet$^{1}$, Donatien Bertschy$^{1}$, Arnaud Ralko$^{1}$, C\'edric Poulain$^{1,2}$ }
\address{$^1$Univ. Grenoble Alpes, CNRS, Grenoble INP, Institut Neel, F-38000 Grenoble, France}
\address{$^2$Univ. Grenoble Alpes, CEA, LETI,  F-38000 Grenoble, France}

\email{cedric.poulain@neel.cnrs.fr}
\begin{abstract}
We present a mechanical analog of a quantum wave-particle duality: 
a vibrating string threaded through 
a freely moving  bead or `masslet'. For small string amplitudes, the particle
movement is governed by a set of non-linear dynamical equations that couple the  wave field to the masslet dynamics. Under specific
conditions, the particle achieves a regime of {\it transparency} in  which the field and the particle's dynamics appear decoupled. In that special case, the particle conserves its momentum and a guiding wave obeying a Klein-Gordon equation, with real or imaginary mass, emerges. Similar to the double-solution theory of de Broglie, this guiding wave is locked in phase with a modulating group-wave co-moving with the particle. Interestingly, both subsonic and supersonic particles can fall into a quantum regime as with the slower-than-light bradyons and hypothetical, faster-than-light tachyons of particle physics.
		
\end{abstract}

\maketitle
\section{Introduction}
The foundations of Quantum mechanics (QM) mainly rely on the pioneering work of
Louis de Broglie elaborated during his PhD \cite{debroglie1925} and for which
he received the Nobel prize.  The key assumption of de Broglie's intuitive
approach is that any mass $m_{\rm p}$ of matter acts like a clock of pulsation
$\omega_{\rm p}$ such that its mass energy  $m_{\rm p}c^2$ balances  its
vibrational energy $\hbar\omega_{\rm p}$ where $c$ is the light velocity and
$\hbar$ the Planck constant.  Then, relying on special relativity to estimate
how this moving clock would appear for an immobile observer, de Broglie showed
that the clock must be in phase with a superluminal phase wave, the guiding
wave, 
giving birth to the celebrated but still mysterious wave-particle duality.
After de Broglie discovered this phase wave, he proposed in late 1926 a
		mechanical analog \cite{debroglie1926,debroglie1927,debroglie1987},
		`the double solution theory'. This analog was subsequently followed by
		some hydrodynamical and very interesting analogs {\it e.g.} the
		Madelung approach~\cite{Madelung1927}. In the same series of works
		\cite{debroglie1927,Valentini}, de Broglie also introduced the
		pilot-wave interpretation  nowadays known  as de Broglie-Bohm -- or
		`Bohmian' mechanics -- after its rediscovery in 1952 by D.
		Bohm~\cite{Bohm1952,Hiley}. 
		While the quest for a classical analog of QM is legitimate, it is far
		from being an easy task and de Broglie failed in extending his earlier
		results. Quite recently, the interest on this subject was renewed by
		the pioneering work initiated by Couder and Fort on bouncing droplets
		\cite{Couder2005,Couder2006,Bush2013} in which one or several droplets
		hit a vertically shaken bath and generate a surface wave. Among other
		works (see \cite{Bush2015} for a review), these droplets, sometimes
		referred to as walkers, were shown to not only mimic a wave particle
		duality at macroscale, but also to reproduce most, if not all, of QM
		features. This is not that surprising because these models share some
		features with the double solution and pilot wave theories
		\cite{debroglie1927,Bohm1952}, known to be possible alternative
		interpretations, however deterministic, of quantum mechanics.

		In this paper, we focus on a type of mechanical analog closer in spirit
		to the original double solution of de Broglie, i.e., transverse waves on which a small bead of mass $m_{\rm
		p}$ and of stiffness $k_{\rm p}$ 
		is submitted to the string  impulse
		  and moves without friction.  This
		`masslet' acts as a particle somewhat similar to a free-to-move defect or
		impedance jump; the density and elasticity are locally altered at  the particle
		location. Its  dynamical behavior is governed by usual momentum transfer
		at the impedance jump, giving rise to reflection, transmission or
		absorption of the incoming wave. The masslet being free to move, these
		mechanisms are accompanied with the radiation force phenomenon common
		to acoustic and electromagnetic fields. This drives the particle in the
		whole field -- incident and scattered -- and eventually leads to memory
		effects. 
		In the spirit of  de Broglie's assumptions, the sliding masslet can 
		be viewed as a   a moving clock of pulsation $\omega_{\rm p}=\sqrt{k_{\rm
		p}/m_{\rm p}}$ in its rest frame.

		In the past, quite similar models
		 were first proposed by Rayleigh and
		Helmholtz~\cite{Rayleigh,Helmholtz,Morse} to study the vibrations of a
		loaded string.  More recently, Boudaoud \textit{et al.} studied the
		self-adaptation of free-to-move beads on a string submitted to acoustic
		noise  \cite{Boudaoud1999}  in the context of soap film and pattern formation. A few years ago, Borghesi  showed
		relying on a relativistic framework, that such a  system yields a wave
		particle duality  governed by the also relativistic Klein-Gordon equation
		\cite{Borghesi2017}. Here, we restrict ourselves to  the non-relativistic Newtonian
		framework and unravel the emergence of a {\it transparency} regime in
		which the particle-string interaction, and thus the radiation force,
		vanish.  As we show, this regime, is reminiscent of de Broglie's double-solution~\cite{debroglie1927}  and 
		leads to a Schr\"odinger equation for the phase wave associated to the particle.
		Very interestingly, two classes of transparent particles are possible candidates:
		(i) the class of subsonic particles that travel uniformly along the string with a velocity smaller than the speed of sound on the string
		and (ii) the class of supersonic (i.~e., faster-than-sound) particles. These two families of dynamical motions are
		hereafter referred to as bradyons and tachyons, respectively in reference 
		to their quantum  counterparts in particle physics (where the speed of sound must be replaced by the speed of light).
		The paper is structured  as follows: In Section.~\ref{sec2}, we describe the system and derive the general dynamical equations. We discuss
		analytical solutions and present the numerical approach employed for
		studying the dynamics. In Section.~\ref{sec3}, we discuss in detail the
		transparency regimes by focusing on the two quantum analogs of
		bradyons and tachyons.
		Concluding remarks are addressed in Section.~\ref{sec4}.  


\section{The string-masslet system}
\label{sec2}
\subsection{General description}
		Our system consists of  an elastic string of linear density $\lambda$
		stretched along $x$  by  tension $T$. The string oscillates in the
		transverse $z$-direction and the vibration is characterized by the field $u(t,x)$. 
The effect of gravity is neglected throughout the paper.
	 A ring-shaped mass $m_p$ -- or bead -- is threaded on this string at $x=x_{\rm p}$
		and can slide without friction.

\begin{figure}
\includegraphics[width=8.5 cm]{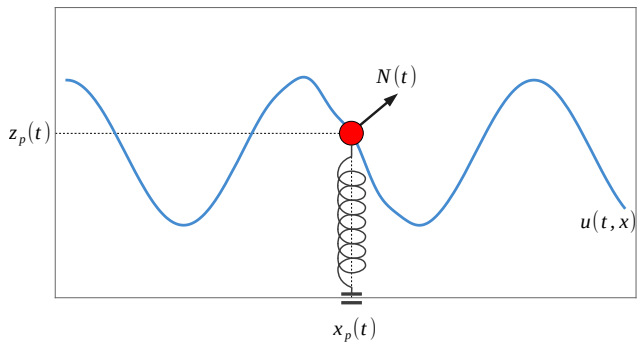} 
\caption{
		Sketch of the string-mass mechanical analog.  A bead of coordinates
		($x_{\rm p},z_{\rm p}$) slides without friction on a string with local
		transverse amplitude $u(t,x)$. A vertical spring acts on the bead as as
		a vertical restoring force and features  an internal clock in the de
		Broglie double solution of quantum mechanics.   } \label{fig:princ}

\end{figure}

Importantly, in order to make this particle behave as a clock, 
the masslet is placed in an additional elastic force field corresponding  to a spring of
stiffness $k_{\rm p}$. Thus, it exerts an elastic force towards the
non-deformed axis of the string ($z=0$) (see illustration in	Fig.~\ref{fig:princ}).

Since the mass cannot escape the string, its vertical position $z_{\rm p}$ must
satisfy the holonomic condition  $z_{\rm p}(t)=u(t,x_{\rm p}(t))$ at any time
$t$.  Furthermore, the absence of any dissipation (and in particular of
friction between the particle and the string) imposes that the reaction force
$N(t)$ from the string to the particle is locally normal to the string at the
particle's location (in the limit of small vibration amplitudes \textit{i.e}
the reaction force is mainly vertical).  When the string vibrates, the mass is
accelerated vertically and horizontally by the local string acceleration and
subsequently moves along it.  Since the particle has locally a density or elasticity which are
different from the string, its inertia will in turn stretch the string and act
as a source for the field $u(t,x)$, generating scattered waves.
Thereby,  the particle acts as a clock moved
by and within the wave field that it has formerly produced.  

 In this work, we only study non-relativistic movements, \textit{i.e.}
we assume that the  velocity of particle and wave are much smaller than the
celerity of light defined by the velocity $c$ of the elastic waves along the
string. This allows us to define  two distinct regimes associated to the Mach
number $Ma=|v_{{\rm p},x}/c|$ where $v_{{\rm p},x}$ is the horizontal velocity
of the particle. $Ma < 1$ is the subsonic regime, while  $Ma >
1$ is the supersonic one.  As we show in the following, the subsonic
regime is physically equivalent to the bradyonic regime as defined in particle
physics for slower-than-light quanta, while the supersonic regime is similar to the tachyonic one with faster than light quantum particles.
The mechanical analogy with special relativity is remarkable and we demonstrate
that the application of an equivalent Lorentz transformation  (where the sound velocity replaces the velocity of light) is crucial for a
quantum description of mechanical analogs.

\subsection{Deriving the equations}

		Since our system possesses several degrees of freedom coupled by a
		holonomic condition, the Lagrangian formalism is well suited to derive
		the equations of motion.
Obviously, the total action $I$ of the system can be split in three parts as
 \begin{equation}
I=\int dt L_{\rm p} +\iint dt dx\mathcal{L}_{\rm s} +\int dt L_{\rm int.}\label{eq:It}
\end{equation} 
	where $L_{\rm p}$ and $\mathcal{L}_{\rm s}$ are respectively the
	Lagrangian of the particle and the Lagrangian density of the string (indicated by the
	cursive letter) in the absence of the holonomic coupling constraint $L_{\rm
	int.}$. We focus first on the expression of the Lagrangian $L_{\rm p}$ of
	the particle in our system. 
	Since we impose an attraction towards the string baseline, a potential 
	\begin{equation}
	V(z_{\rm p})=\half m_{\rm p}\omega_{\rm p}^2 z_{\rm p}^2 \label{eq:harmPot}
	\end{equation}
	is introduced, with $\omega_{\rm p}$ a pulsation or equivalently $k_{\rm
	p}=m_{\rm p}\omega_{\rm p}^2$ a stiffness applied to the particle solely.
	The particle Lagrangian then takes the form 
	\begin{equation}
			L_{\rm p}(t,z_{\rm p},v_{{\rm	p}})=\half m_{\rm p}v_{{\rm p}}^2-V(z_{\rm p})\label{eq:Lp}
	\end{equation} 
	where we have introduced $\vec{v}_p = v_{p,x} \vec{u}_x + v_{p,z} \vec{u}_z$ with
	$v_{{\rm p},x}(t)=\frac{dx_{\rm p}(t)}{dt}$ and $v_{{\rm p},z}(t)=\frac{d
	z_{\rm p}(t)}{d t}$ the longitudinal and transverse particle velocity
	respectively.  Without the holonomic constraint, the variational Lagrange
	principle $\delta[\int dt L_{\rm p}]=0$ leads to the usual Newtonian
	equations 
\begin{eqnarray}
m_{\rm p}\ddot{x}_{\rm p}(t)=0, & m_{\rm p}[\ddot{z}_{\rm p}(t)+\omega_{\rm p}^2 z_{\rm p}(t)]=0.\label{eq:N}
\end{eqnarray}
Similarly, $\mathcal{L}_{\rm s}$ is obtained by taking the continuum limit of a
chain of springs with constant stiffness and considering small amplitude vibrations (i.e., neglecting nonlinearities): 
\begin{equation}
\mathcal{L}_{\rm s}(u, \partial_t u, \partial_x u)=\half\lambda(\partial_t u(t,x))^2-\half T(\partial_x u(t,x))^2\label{eq:Ls}
\end{equation}
In this equation, $\lambda$ and $T$ 
are respectively the linear density and the tension of the string. We also
define our local speed of sound $c = \sqrt{T/\lambda}$ of the transverse
waves propagating along  the string.
Once again, for $L_{\rm int.} = 0$, one can obtain from the variational principle $\delta[\int dt\int dx\mathcal{L}_{\rm s}]=0$  the wave-equation:
\begin{equation}
\lambda\partial^2_t u(t,x) - T\partial^2_x u(t,x) =  \square u(t,x) = 0\label{eq:field}
\end{equation} 
where $\square$ is the usual linear d'Alembert operator.

The specific nature of the holonomic constraint is ideally grasped by the coupling Lagrangian:
 \begin{equation}
L_{\rm int.}(t,x_{\rm p},z_{\rm p},N)=N\cdot[z_{\rm p}-u(t,x_{\rm p})],\label{eq:Lint}
\end{equation} or equivalently by a Lagrangian density  $\mathcal{L}_{\rm int.}$ (such that $L_{\rm int.}=\int dx\mathcal{L}_{\rm int.} $):
\begin{equation}
\mathcal{L}_{\rm int.}(u,x_{\rm p},z_{\rm p},N)=N\cdot[z_{\rm p}-u(t,x)]\delta(x-x_{\rm p}),\label{eq:Ldenint}
\end{equation}
 where $N(t)$ is a Lagrange multiplier defining an additional variable in the variational problem. Variations with respect to $N$ lead automatically to the holonomic constraint:
\begin{equation}
z_{\rm p}=u(t,x_{\rm p}).\label{eq:holo}
\end{equation} 
Taking into account the action of the mass on the string, the wave dynamics follows a new equation obtained from the variational principle $\delta[\int dt\int dx(\mathcal{L}_{\rm s}+\mathcal{L}_{\rm int.})]=0$, whereas symmetrically the modified dynamic of the sliding mass is obtained from $\delta[\int dt (L_{\rm p}+L_{\rm int.})]=0$.
Altogether this leads to the following set of coupled equations:  
\begin{subequations}
\begin{eqnarray}
m_{\rm p}\ddot{x}_{\rm p}(t)=-\partial_{x} u(t,x)|_{x=x_{\rm p}(t)}\, N(t),\label{total1}\\
N(t)=m_{\rm p} \left[\ddot{z}_{\rm p}(t)+\omega_{\rm p}^2 z_{\rm p}(t)\right],\label{total2}\\
\square u(t,x)=-\frac{N(t)}{T}\delta(x-x_{\rm p}(t))\label{total3} 
\end{eqnarray}
\end{subequations}
 with the holonomic condition Eq.~\ref{eq:holo}. Interestingly, after elimination of $N(t)$ Eqs.~\ref{total1} and \ref{total2} can equivalently be rewritten as 
\begin{eqnarray}
\ddot{x}_{\rm p}(t)=-\partial_{x} u(t,x)|_{x=x_{\rm p}(t)}\left[\ddot{z}_{\rm p}(t)+\omega_{\rm p}^2 z_{\rm p}(t)\right]\label{total4} 
\end{eqnarray}
Remarkably, this equation is  independent of the mass  $m_{\rm p}$ a fact which is reminiscent of acoustic analogs of gravitational forces in hydrodynamical systems~\cite{Duplat}. We also point-out that Eqs.~\ref{total1}, \ref{total2}, and \ref{total3} can easily be interpreted using a Newtonian language. The Lagrange multiplier $N(t)$ is the vertical reaction force acting on the sliding mass while $-\partial_{x} u(t,x)|_{x=x_{\rm p}(t)}\, N(t)$ is the horizontal component of this reaction force (i.e., in the linear limit of small wave-amplitude we have $\partial_{x} u(t,x)|_{x=x_{\rm p}(t)}=\tan{\theta}\simeq \sin{\theta} $ where $\theta$ is the angle between the reaction force $N$ and the $z$ vertical direction). We emphasize that Eq.~\ref{total3} can be rewritten as $\tilde{\lambda}(t,x)\partial^2_t u(t,x)=T\partial^2_x u(t,x)=0$ 
 with $\tilde{\lambda}(t,x)=\lambda+ m\delta(x-x_{\rm p}(t))$. In other words  the mass $m$ can also be interpreted as a local  translatable defect in the linear density $\lambda$. This could have an impact for physical interpretations.  Moreover, the  previous dynamics is completed by an analysis of energy and momentum conservation in the coupled system (\textit{i.e.}, of prime integrals of motion).  From  the full action and Lagrangian Eq.~\ref{eq:It} (or alternatively from Eqs.~\ref{total1}, \ref{total2}, and \ref{total3}) we deduce after lengthy but straightforward calculations the  local energy-momentum conservation laws for the field coupled to the mass:
\begin{eqnarray}
\partial_t\varepsilon(t,x)+\partial_xS_x(t,x)=-N\delta(x-x_{\rm p}(t)\partial_{t} u(t,x)|_{x=x_{\rm p}(t)}\nonumber\\
\partial_tg_x(t,x)+\partial_xT_{xx}(t,x)=N\delta(x-x_{\rm p}(t)\partial_{x} u(t,x)|_{x=x_{\rm p}(t)}\nonumber\\
\label{conservation} 
\end{eqnarray}
where $\varepsilon=\half T[\frac{1}{c^2}(\partial_t u)^2+(\partial_x u)^2]$ and $g_x=-\frac{T}{c^2}\partial_t u\partial_x u$ are respectively the $u-$field energy and linear (pseudo) momentum density of the acoustic field along the string.  Similarly $S_x=c^2g_x$ and $T_{xx}=\varepsilon$ are respectively the energy and (pseudo) momentum density flow along the $x$ direction ($T_{xx}$ is the constraint tensor of the field  which has only one component here). Moreover, combining Eq.~\ref{conservation}  with  Eqs.~\ref{total1}, \ref{total2}, and \ref{total3} leads to 
 \begin{subequations}
\begin{eqnarray}
\frac{d}{dt}\left\{\int dx\varepsilon(t,x)+\half m_{\rm p}(\dot{x}_{\rm p})^2\right.\nonumber\\
+\left.\half m_{\rm p} \left[(\dot{z}_{\rm p})^2+\omega_{\rm p}^2 (z_{\rm p})^2\right]\right\}=0\\
\frac{d}{dt}\left\{\int dxg_x(t,x)+ m_{\rm p}\dot{x}_{\rm p}\right\}=0
\end{eqnarray}
\label{conservationb}
\end{subequations}  which shows that the total energy and linear momentum of the system are conserved.\\ 
\indent We stress  that despite its non-relativistic nature our model differs from the one proposed in \cite{Borghesi2017} which considered a Klein-Gordon wave equation from the start, \textit{i.e.}, $[\square +\frac{\Omega_m^2}{c^2}]u(t,x)=-\frac{N(t)}{T}\delta(x-x_{\rm p}(t))$.  In \cite{Borghesi2017} it was shown that such a Klein-Gordon equation with a source term can be used to generate a mechanical analog of  wave-particle duality. Here we show that it is not necessary to introduce such a complication.  An inhomogeneous d'Alembert equation, \textit{i.e.}, as in Eq.~\ref{total3}, is already sufficient to reproduce wave-particle duality. As we show in the following this is in complete agreement with original ideas developed by de Broglie before 1927 \cite{debroglie1925b}.     
\section{The transparency regime}
\label{sec3}
\subsection{The bradyonic or subsonic  regime: wave-particle duality }
Let us now focus on a particular class of motion for which the coupling force $N(t)$ between the field and the particle cancels:
\begin{equation}
m_{\rm p}[\ddot{z}_{\rm p}(t)+\omega_{\rm p}^2 z_{\rm p}(t)]=N(t)=0\label{eq:Ntransp} \quad{\rm(transparency)}
\end{equation}
yielding a purely oscillatory motion
\begin{equation}
z_{\rm p}(t)=A\cos{(\omega_{\rm p}t+\varphi)}\label{eq:zHarm}
\end{equation} 
with $A$ and $\varphi$ real constants.
Eqs.~\ref{total1} and \ref{total3} yield:
\begin{eqnarray}
\ddot{x}_{\rm p}(t)=0, &&
\square u(t,x)=0\label{eq:uTransp}
\end{eqnarray} 
which results in an inertial movement for the particle:
\begin{equation}
x_{\rm p}(t)=v_{\rm p}t+x_{\rm p,i}\label{eq:xUnif}
\end{equation}
with $x_{\rm p}(0)=x_{\rm p,i}$ and $v_{\rm p}(t)=v_{\rm p}$ two real constants.  In this configuration the different degrees of freedom are decoupled and in particular $\int dx\varepsilon(t,x)$, $\half m_{\rm p}(\dot{x}_{\rm p})^2$, $\half m_{\rm p} \left[(\dot{z}_{\rm p})^2+\omega_{\rm p}^2 (z_{\rm p})^2\right]$, $\int dx g_x(t,x)$ and $m_{\rm p}\dot{x}_{\rm p}$ are constant of motions. We emphasize that the oscillatory motion $z_{\rm p}(t)$ is reminiscent of de Broglie's original idea \cite{debroglie1925,debroglie1925b} of associating a local clock to any particle.  Here, we have a more detailed  mechanical model for which we see that the uniform motion given by Eq.~\ref{eq:xUnif} is dynamically linked to the harmonic oscillation defined by Eq.~\ref{eq:zHarm}.\\
\indent Concerning the string field $u(t,x)$, several solutions are \textit{a priori} available.  Indeed the general solution  of $\square u(t,x)=0$ reads $u(t,x)=f(t-x/c)+g(t+x/c)$ corresponding to two arbitrary pulses $f(t)$ and $g(t)$ propagating respectively along the $+x$ and $-x$ direction. Considering for example the case $g=0$ and using the holonomic condition  Eq.~\ref{eq:holo} one gets  $f(t(1-\frac{v_{\rm p}}{c})-\frac{x_{\rm p,i}}{c})=A\cos{(\omega_{\rm p}t+\varphi)}$, i.e., 
\begin{eqnarray}
f(t)=A\cos{\left[\frac{\omega_{\rm p}}{1-\frac{v_{\rm p}}{c}}\left(t+\frac{x_{\rm p,i}}{c}\right)+\varphi\right]}.\label{eq:surfer}
\end{eqnarray}    
This motion would correspond to a mass `surfing' on a monochromatic propagative wave with pulsation $\frac{\omega_{\rm p}}{1-\frac{v_{\rm p}}{c}}$.  However, while Eq.~\ref{eq:surfer} is interesting in itself we are not going here to further develop this approach. Instead, we now follow the physical intuitions of de Broglie and seek solutions
for the homogeneous equation $\square u=0$ such that 
in a co-moving inertial frame $\mathcal{R}'$ translating at velocity $v_{\rm p}$
with the particle, the field $u$ would appear as stationary.
Interestingly, a Galilean transformation of the coordinates ($t'=t$, $x'=x-v_{\rm p}t$) which lets the time flow unchanged 
fails to bring a standing solution.
This coordinate change, although legitimate here for a Newtonian approach (small Mach number)
is however too crude to allow for clock synchronization at a distance.
A  coordinate transformation that  does not let the time unaltered
is required to bring a stationary solution. Actually, a `first-order' Poincar\'e-Lorentz transformation of the form:  $x'=x-v_{\rm p}t$, $t'=t-\frac{v_{\rm p}}{c^2}x$ is already sufficient to ensure the invariance of the free wave equation. This is because  the term $-\frac{v_{\rm p}}{c^2}x$ results from  a time synchronization procedure for clocks located at different points of inertial frames $\mathcal{R}$ and $\mathcal{R}'$  as already proposed by Poincar\'e in 1900~\cite{Poincare}. 
 Moreover, the d'Alembert operator $\square$ being invariant 
under the usual Lorentz transformation $\square=\square'$, we make use of the following coordinate change:
\begin{equation}
x'=\gamma_{\rm p}\left(x-v_{\rm p}t\right),\quad t'=\gamma_{\rm p}\left(t-\frac{v_{\rm p}}{c^2}x\right), \label{eq:lorTransf} 
\end{equation}  with $\gamma_{\rm p}^{-1} = \sqrt{1-\frac{v_{\rm p}^2}{c^2}}$ 
so that in the Lorenz-Poincar\'e group in dimension 1+1,
the field $u(t,x)$ appears as a scalar invariant field, \textit{i.e.}, $u(t,x)=u'(t',x')$. We point out that the variable $t'$ and $x'$ have no direct physical meaning here since we are working in the context of Newtonian dynamics where the time is absolute. However, Eq.~\ref{eq:lorTransf} is used as a mathematical tool for finding the solutions of the d'Alembert equation under the restriction $Ma<1$ (in this context we emphasize that Voigt~\cite{voigt} already used the Lorentz transformation as a mathematical tool in optics but devoid of physical interpretation).   
Searching for a standing solution in the Lorentzian co-moving frame $\mathcal{R}'$
that ensures a `time-space' separation $u'(t',x')=F(t')G(x')$, one gets:
 
\begin{equation}
\frac{1}{c^2}\frac{d^2F}{dt'^2}G-F\frac{d^2G}{dx'^2}=0
\end{equation}
which turns into a set of equation:
\begin{eqnarray}
\frac{d^2F}{dt'^2}+\omega'^2F=0, \quad
\frac{d^2G}{dx'^2}+\frac{\omega'^2}{c^2}G=0
\end{eqnarray} 
with $\omega'$ a complex constant to be determined.
For the case of interest when $\omega^\prime$ is real, $F$ and $G$ are harmonic so that we  obtain an `amplitude modulated' field $u'(t',x')$:
\begin{eqnarray}
u'(t',x')=B\cos{\left(\omega't'+\eta\right)}\cos{\left(\frac{\omega'}{c}x'+\xi\right)}=u(t,x)\nonumber\\=B\cos{\left[\omega'\gamma_{\rm p}\left(t-\frac{v_{\rm p}}{c^2}x\right)+\eta\right]}\cos{\left[\frac{\omega'}{c}\gamma_{\rm p}\left(x-v_{\rm p}t\right)+\xi\right]}\nonumber\\ \label{eq:uExpr}
\end{eqnarray}
 $B$, $\eta$ and $\xi$ being three real constants.
   They can be determined by using the holonomic condition expressed for the case of the  uniform motion:
$z_{\rm p}(t)=u(t,x=v_{\rm p}t+x_{\rm p,i})$.
It follows that:
\begin{eqnarray}
u(t,x=vt+x_{\rm p,i})=&B&\cos{\left(\frac{\omega'}{\gamma_{\rm p}}t+\eta-\frac{\omega'\gamma_{\rm p} x_{\rm p,i}v_{\rm p}}{c^2}\right)}\nonumber\\ &\times &\cos{\left(\frac{\omega'}{c}\gamma_{\rm p}x_{\rm p,i}+\xi\right)}\nonumber
\end{eqnarray} 
which, by identification with 
$z_0(t)$ given by Eq.~\ref{eq:zHarm} yields:
\begin{eqnarray}
\varphi = \eta - \frac{\omega'\gamma_{\rm p} x_{\rm p,i}v_{\rm p}}{c^2}+ 2\pi n & (\textrm{with $n\in \mathbb{Z}$}),\label{eq:phiCond}
\end{eqnarray}
\begin{equation} 
A=B\cos{\left(\frac{\omega'}{c}\gamma_{\rm p}x_{\rm p,i}+\xi\right)},\label{eq:ACond}
\end{equation} 
and above all the relation:
\begin{equation}
\omega_{\rm p}=\frac{\omega'}{\gamma_{\rm p}}.\label{eq:clocks}
\end{equation} 

To give a meaningful interpretation of this condition,
we can write: $u'=u$ in Eq.~\ref{eq:uExpr} in the form  $u=B\cos{S_{\textrm{brad.}}}\cos{\Phi_{\textrm{brad.}}}$ with
\begin{eqnarray}
S_{\textrm{brad.}}=\omega't'+\eta=\omega t-k x+\eta\nonumber\\
\Phi_{\textrm{brad.}}=\frac{\omega'}{c}x'+\xi=\frac{\omega}{c}(x-v_{\rm p}t)+\xi\label{eq:phases}
\end{eqnarray}
with $\omega=\omega'\gamma_{\rm p}$, $k=\omega v_{\rm p}/c^2$. 
Besides, the following dispersion relation applies
\begin{equation}
\frac{\omega^2}{c^2}-k^2=\frac{\omega'^2}{c^2}\label{eq:disp}
\end{equation}
between the pulsation $\omega$ and the wave-vector $k$. 
The quantity  $S_{\textrm{brad.}}$ plays the role of a phase for a plane (carrying) wave, 
$\psi(t,x)=e^{iS_{\textrm{brad.}}}$ solution of the Klein-Gordon equation:
\begin{equation}
\left(\frac{1}{c^2}\partial^2_t -\partial^2_x\right)\psi(t,x)=-\frac{\omega'^2}{c^2}\psi(t,x).\label{eq:KG}
\end{equation}
The phase velocity $v_{\textrm{ph.}}$ is defined by the condition $dS_{\textrm{brad.}} = 0$, which implies
\begin{equation}
v_{\textrm{ph.}}=\frac{\omega}{k}=\frac{c^2}{v_{\rm p}}>c\label{eq:phaseVelocity}
\end{equation} 
in accordance with the formulas obtained by Louis de Broglie in his PhD manuscript \cite{debroglie1926}.
Besides, $\Phi_{\textrm{brad.}}$ in Eq.~\ref{eq:phases} defines 
an envelop  (\textit{i.e.}, a group) velocity which identifies with the particle's velocity $v_{\rm p}$ (setting $d\Phi_{\textrm{brad.}}=0$ we get $dx/dt=v_{\rm p}$). Furthermore, we deduce the Rayleigh formula $v_{\textrm{gr.}}=\frac{d\omega}{dk}=v_{\rm p}$ which was also obtained by de Broglie. This is clearly reminiscent of Hamilton formula $\frac{dH_{\rm p}}{dP_{\rm p}}=v_{\rm p}$ if we identify the Hamiltonian or energy function $H_{\rm p}:=m_{\rm p}c^2\gamma_{\rm p}$ and the linear momentum $P_{\rm p}:=m_{\rm p}\gamma_{\rm p}v_{\rm p}$ with respectively $Q\omega$ and $Qk$ where $Q$ is a constant having the dimension of an action.  The similarity between  $Q$ and $\hbar$ is clear and is reenforced if we write  Eq.~\ref{eq:KG} in the limit $M_a\ll 1$ as 
 \begin{equation}
iQ\partial_t\Psi\simeq -\frac{Q^2}{2m_{\rm p}}\partial^2_x\psi(t,x)+m_{\rm p}c^2\psi(t,x)\label{eq:Sch}
\end{equation}  
which is identical to Schr\"odinger's equation after the substitution $Q\rightarrow\hbar$ (similarly in the Klein-Gordon Eq.~\ref{eq:KG} we can replace $\frac{\omega'^2}{c^2}$ by $\frac{m_{\rm p}^2c^2}{Q^2}$in agreement with standard text-books). Moreover, we can also write $H_{\rm p}=-Q\partial_t S_{\textrm{brad.}}$ and $P_{\rm p}=Q\partial_x S_{\textrm{brad.}}$ which are reminiscent of Hamilton-Jacobi equations with $QS$ playing the role of an action and therefore we have 
\begin{equation}
v_{\rm p}=-c^2\frac{\partial_x S_{\textrm{brad.}}}{\partial_t S_{\textrm{brad.}}}
\label{eq:Bohm}
\end{equation}  which is the guidance formula introduced by de Broglie in his pilot-wave interpretation~\cite{debroglie1927,Valentini} and which leads to $v_{\rm p}=Q\frac{\partial_x S_{\textrm{brad.}}}{m_{\rm p}}$ in the limit $Ma\ll 1$ in agreement with Bohmian mechanics~\cite{Valentini,Bohm1952}.
 Therefore, altogether we recover de Broglie's assumptions casting the 
double solution theory \cite{debroglie1927} with 
a $\psi$-wave also called the guiding wave (solution of the Klein-Gordon equation)
and the $u$-wave (solution of the homogeneous d'Alembert equation)
associated with the particle's movement. Importantly, in our approach (and in contradistinction to \cite{Borghesi2017}) we start from the d'Alembert equation and not from the Klein-Gordon equation. Still, we are able to obtain a guiding-wave $\psi$ which is solution of Eq.~\ref{eq:KG} \textit{i.e.}, the Klein-Gordon equation. In other words the mass term of the Klein-Gordon equation has been generated from the $u-$field itself. This agrees with a model already presented by de Broglie in 1925 \cite{debroglie1925b}, \textit{i.e.}, two years before the model named traditionally the `double solution'\cite{debroglie1927} and based on the Klein-Gordon equation.     

\begin{figure}
\includegraphics[width=\linewidth]{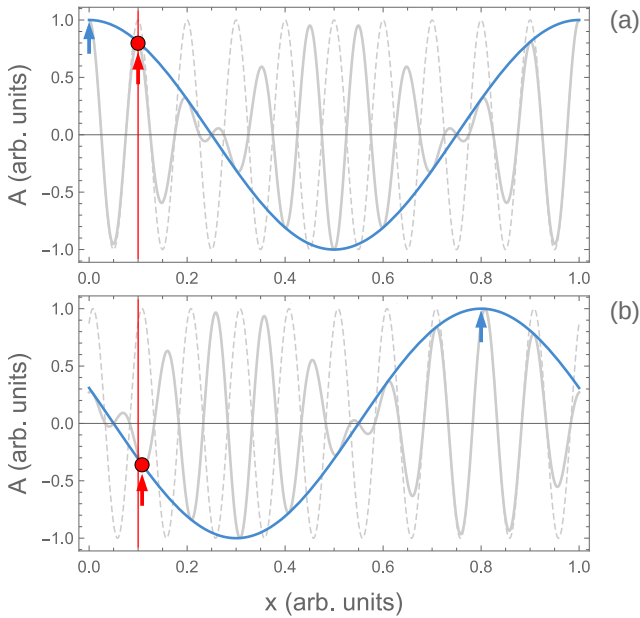} 
\caption{The transparent regime for a bradyonic (\textit{i.e.}, subsonic) particle
plotted from the analytical solutions described in the paper. 
An equivalent  animation obtained from the numerical code of the full set of equation
is available in the supplemental material~\cite{Movie}. 
For this example, we have $v_{\rm p}/c=0.1$, $x_{\rm p,i}=0.1$ for a string length of $L=1$ and $c=1$ (in natural units). We have chosen a $\psi-$wave temporal period $T_{\textrm{phase}}=\frac{2\pi}{\omega}=10$ and $\eta=\xi=0$.  
The continuous thick gray line is the total field $u(x,t)$ 
which can be decomposed into the phase-wave $\psi$ (continuous blue line)
and the group wave (dashed gray line).
The whole field is shown at two different instants (a) $t=0$ and  (b) $t=0.08$ clearly demonstrating the
subsonic (or bradyonic) regime of the particle.
The particle's initial condition is set so that it is at a maximum of the group phase $\Phi_{\textrm{brad.}}$ (see Eq.~\ref{eq:phases})
and is clearly locked to it along time. The red arrows in panel (a) and (b) compare the position of the particle at time $t=0$ and $t=0.08$ whereas the blue arrow show the position  and displacement of a phase maximum at the two instants $t=0$ and $t=0.08$.   
}
\label{fig:evo1}
\end{figure}

The condition (\ref{eq:clocks}) on the frequency can also be written in terms of phase and becomes
a phase-locking condition, since for 
$x=x_{\rm p}(t)$ one gets:
\begin{equation}
S_{\textrm{brad.}}=\omega't'+\eta=\frac{\omega'}{\gamma_{\rm p}}t+\varphi=\omega_{\rm p}t+\varphi\label{eq:phaseLock}
\end{equation}
which expresses the phase-locking of the particle's clock  (with pulsation $\omega_{\rm p}$)
to  that of the wave (with pulsation $\omega$ or $\omega'$).
Following de Broglie \cite{debroglie1925b,debroglie1925c},
we can rewrite the total wave $u$ as a sum  of waves (by means of  the trigonometric identity: $2\cos{F}\cos{G}=\cos{(F+G)}+\cos{(F-G)}$) to 
give a physically meaningful interpretation

\begin{eqnarray}
u'(t',x')=\frac{B}{2}\cos{\left[\omega'\left(t'+\frac{x'}{c}\right)+\eta+\xi\right]}\nonumber\\+\frac{B}{2}\cos{\left[\omega'\left(t'-\frac{x'}{c}\right)+\eta-\xi\right]}\label{eq:uPMprime}
\end{eqnarray} or equivalently 
\begin{eqnarray}
u(t,x)&=&\frac{B}{2}\cos{\left[\omega'\gamma_{\rm p}(1-\frac{v_{\rm p}}{c})(t+\frac{x}{c})+\eta+\xi\right]}\nonumber\\
&+&\frac{B}{2}\cos{\left[\omega'\gamma_{\rm p}(1+\frac{v_{\rm p}}{c})(t-\frac{x}{c})+\eta-\xi\right]}\label{eq:uPM}
\end{eqnarray} 
It is another way to express $u$ as the sum of two counter propagating waves $u(t,x)=u_+(t,x)+u_-(t,x)$ with 
(i) $u_-=\frac{B}{2}\cos{\left[\omega_-t+\omega_-x/c+\eta+\xi\right]}$ a wave with a low-frequency Doppler shift $\omega_-=\omega(1-\frac{v_{\rm p}}{c})$ propagating along the $-x$ direction,
and (ii) $u_+=\frac{B}{2}\cos{\left[\omega_+t-\omega_+x/c+\eta-\xi\right]}$ a wave with a high-frequency Doppler shift $\omega_+=\omega(1+\frac{v_{\rm p}}{c})$ propagating along the $+x$ direction.
Experimentally, this decomposition will allow us  to generate 
the resulting  modulated $u$ wave appearing in Eq.~\ref{eq:uExpr} as the sum of two plane waves.\\

The analytical solution is plotted in Fig.~\ref{fig:evo1} for this transparency regime (the normalized parameters are indicated in the caption).  Going back to Eq.~\ref{eq:phases} we have for the phase-wave a temporal period $T_{\textrm{phase}}=\frac{2\pi}{\omega}$ and a spatial period $\lambda_{\textrm{phase}}=\frac{2\pi}{\omega}\frac{c^2}{v_{\rm p}}$ which have to be compared with the temporal and spatial period of the group wave  $T_{\textrm{group}}=\frac{2\pi}{\omega}\frac{c}{v_{\rm p}}\geq T_{\textrm{phase}}$, $\lambda_{\textrm{group}}=\frac{2\pi}{\omega}c\leq \lambda_{\textrm{phase}}$ (as illustrated in Fig.~\ref{fig:evo1} for $\lambda_{\textrm{phase}}/\lambda_{\textrm{group}}=10$. Importantly, we have  here the well-known de Broglie formula \cite{debroglie1925}:
 \begin{eqnarray}
P_{\rm p}=\frac{2\pi Q}{\lambda_{\textrm{phase}}}
\end{eqnarray}
\indent It is interesting to note that in de Broglie approach the wave field  was a solution of a non-homogeneous equation  including a source term for the particle $\square u(t,x)\propto\delta(x-x_{\rm p}(t))$ (i.e., like in Eq.~\ref{total3}). Here, instead the transparency condition $N(t)=0$ imposes the field equation $\square u(t,x)=0$ and this simplifies the interpretation of Eq.~\ref{eq:uPM} as the sum of two plane waves whereas de Broglie had to involve a complicated sum of advanced and retarded waves emitted by the particle itself~\cite{debroglie1925b,debroglie1925c}.\\
\indent More generally, it is clear that the equations of motion presented here are strongly non-linear, and except for a  few specific regimes as the ones discussed in the present paper, no analytical solutions are usually reachable. In order to further investigate the physics of our model, we have employed a numerical scheme based on standard finite differences through a Runge-Kutta of order 4 (see for example~\cite{Loustau}). Since the mutual interaction between the mass and the string only enters as an external source term for the each other, we have implemented separately their differential equations and solve the full system self consistently. We first start by fixing initial compatible configurations for both the string and the mass. During the time resolution process, an update of the string is performed, accounting the presence of the mass. Once this space-time update is achieved, we proceed to the update of the mass now accounting back to the new state of the string. And we continue so on and so forth until desired time periods. The convergence is ensured by respecting the von Neumann stability criterion. Despite the above mentioned non-linearities in the complete set of  equations, the algorithm appears to be well stable and reproduces very precisely the analytical solutions discussed here. (transparency). This validates our approach and indicates that the coupling between the mass and the string does not seem to be that crucial in the convergence. For this first work, we have employed our algorithm in order to complement with the dynamics the discussed analytical solutions. In particular, we have recorded a movie~\cite{Movie} showing the behavior of the mass in the transparency regime. The question of other possible emerging exotic regimes in our coupled system is very interesting though, and deserves a proper study. We dedicate a more systematic numerical analysis to a future work.

The calculations above deserve some comments in relation with de Broglie's picture.
Indeed, it is noteworthy that the pulsation  $\omega_{\rm p}$ remains unchanged regardless of the particle's speed
and is in particular different from $\omega^\prime$ (which is velocity dependent).
It is slightly different from de Broglie's results and is a consequence of 
our mixed approach, combining a Lorenz transformation (by means of $c$ the speed of sound) 
in a Newtonian framework for which relativistic 
dynamics do not apply. More precisely, in de Broglie's approach which is fully relativistic  (\textit{i.e.}, with the sound celerity replaced everywhere by the light velocity) the particle's internal clock with pulsation $\omega_{\rm p}$ is defined in the rest-frame $\mathcal{R}'$, not in the laboratory frame $\mathcal{R}$. The phase-locking condition reads now:    
\begin{equation}
S_{\textrm{brad.}}=\omega't'+\eta=\frac{\omega'}{\gamma_{\rm p}}t+\eta-\frac{\omega'\gamma_{\rm p} x_{\rm p,i}v_{\rm p}}{c^2}=\omega_{\rm p}t'+\varphi\label{eq:phaseLockdeb}
\end{equation}
and we have thus $\omega'=\omega_{\rm p}=\frac{\omega}{\gamma_{\rm p}}$ which differs from Eq.~\ref{eq:clocks} by a prefactor $\gamma_{\rm p}$ associated with a relativistic time-dilation (we have also $\eta=\varphi$ since $\varphi$ is now defined in the rest frame\cite{debroglie1925b,Borghesi2017}).  In the non-relativistic limit where  $\frac{v_{\rm p}}{c}\ll 1$  and  $\gamma_{\rm p}\simeq 1$ de Broglie's theory reduces to $\omega\simeq \omega'=\omega_{\rm p}$. This is identical  in our model to  the regime $Ma \ll 1$ (where $c$ is now the sound velocity) so that the difference between the two approaches vanishes for sufficiently slow particle motions. It is interesting to remark  that in the $Ma \ll 1$ regime  the   Hamiltonian $H_{\rm p}:=m_{\rm p}c^2\gamma_{\rm p}$ which was formally introduced reduces to $H_{\rm p}\simeq m_{\rm p}c^2+ \half m_{\rm p}v^2_{\rm p}$  which, up to an additive constant, is the translational kinetic energy associated  with the particle motion along $x$ (similarly $P_{\rm p}\simeq m_{\rm p}v_{\rm p}$ which identifies with the translational linear momentum of the particle). Since $c$ is here the sound velocity, $m_{\rm p}c^2$ can not physically be identified with a rest energy which is a relativistic Einsteinian concept. This once again stresses the similarities and differences between our mechanical analog and de Broglie's own approach. For similar reasons we have no right to identify the integration factor $Q$ discussed previously with the Planck constant $\hbar$. This could only hold in de Broglie's model for a genuine quantum particle. Still, the mechanical analogy works fine for $Ma \ll 1$ and could be actually extended to the relativistic regime by implementing a covariant, \textit{i.e.} Einsteinian, mechanical model. Here, we nevertheless stick to the Newtonian framework which is closer to the experimental realization and is already sufficient to grasp the essential features of de Broglie's mechanical model.    
\subsection{The Tachyonic  or supersonic regime}

So far, we have discussed the dynamics of a particle moving on the string with a velocity $v_{\rm p}$ smaller than that of the waves on this string. Since this motion is completely governed by the interaction with the field $u$, we do not expect the particle to spontaneously cross the sound barrier without the use of an external force. However, one could choose initial conditions such that $v_{\rm p}> c$. We would then find a whole new supersonic regime, which we can investigate. As explained in the introduction we choose to refer to such a particle as a `tachyon' by analogy with the eponymous hypothetical particles introduced in special relativity, which are a class of solution for the dynamics associated with faster than light motions~\cite{Terletsky1,Bilaniuk,Feinberg,Terletsky2} (symmetrically the case studied previously with $v_{\rm p}<c$ is referred to as `bradyonic' motion~\cite{Terletsky1,Bilaniuk,Feinberg,Terletsky2}). Indeed, this velocity $c$ appears in the Lorentz transformation that we use for the field, 
as in  the Lorentz transformation of special relativity, and appears as an asymptotic limit for the velocity $v_{\rm p}$ in both cases.
 However one should once again not confuse those two velocities: it is a fundamental limit within the  special relativity framework, while it is not in the context of a particle sliding on a string.
 Supersonic particles  (analogous to  supersonic aircrafts in a fluid) are indeed possible solutions.
 Thus, it is completely reasonable to study these `acoustic' tachyons, which are  physically viable solutions of the dynamical equations.
\begin{figure}
\includegraphics[width=\linewidth]{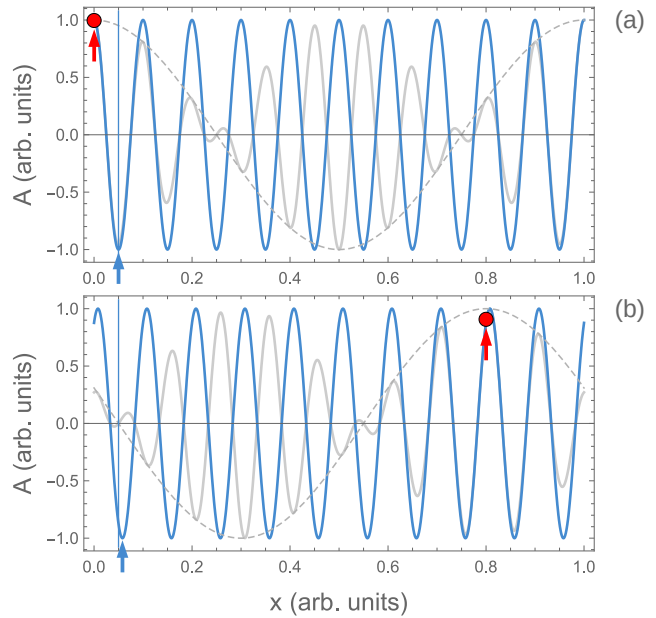} 
\caption{The transparent regime for a tachyonic (i.e., supersonic) particle of velocity $w_{\rm p}/c=10$, and initial coordinate $X_{\rm p,i}=0$ (in natural units). The two panels correspond  to observation times (a) $t=0$ and  (b) $t=0.08$. The other parameters of the wave and string are unchanged with respect to Fig.~\ref{fig:evo1}. In particular, we have $T_{\textrm{phase}}=\frac{2\pi}{\Omega}=\frac{2\pi}{\omega}\frac{c}{v_{\rm p}}=1$ (see text).  Like the subsonic case, the particle is locked to the group wave  $\cos{\Phi_{\textrm{tach.}}}$ (dashed gray line) and is clearly faster than the velocity of the phase-wave $\cos{S_{\textrm{tach.}}}$ on the string (blue continuous line). The red and  blue arrows compare the displacement of the particle and phase wave between  $t=0$ and  $t=0.08$.
}
\label{fig:evo2}
\end{figure}
For this purpose, we see from Eq.~\ref{eq:lorTransf} that if we set $x^\prime = 0$, we get $x = v_{\rm p}t$ and therefore the $ct^\prime$ axis corresponds to the trajectory of a particle with velocity $v_{\rm p}<c$ (this is of course  the definition of a comoving rest frame for such a `bradyonic' motion). If we now set $t^\prime = 0$, we get $x = \frac{c^2}{v_{\rm p}}t = w_{\rm p}t$ which corresponds to a case  where the $x^\prime$ axis identifies with the trajectory of a tachyonic particle with velocity $w_{\rm p} = \frac{c^2}{v_{\rm p}}> c$. This is the first hint that the dynamics of tachyons is completely symmetrical with that of `normal' bradyonic particles. Indeed, we can rewrite Eq.~\ref{eq:lorTransf} as
\begin{equation}
t^\prime = \frac{1}{c}\Gamma_{\rm p}(w_{\rm p}t - x),\quad x^\prime = c\Gamma_{\rm p}\left(\frac{w_{\rm p}}{c^2}x - t\right), \label{eq:tachLorTransf}
\end{equation}  with  $\Gamma_{\rm p} = (\frac{w_{\rm p}^2}{c^2} - 1)^{-1/2}=\gamma_{\rm p}\frac{v_{\rm p}}{c} $ and  where we see, by comparing with the original Lorentz transformation, that the roles of space and time have been reversed in the tachyonic case, compared to the bradyonic one. Furthermore, like for Eq.~\ref{eq:lorTransf} the physical meaning of Eq.~\ref{eq:tachLorTransf} is not immediate and here we use it mainly as a mathematical tool for guessing at an interesting solution of the wave equation. More precisely, let us consider once again the stationary field 
\begin{equation}
u^\prime(t^\prime,x^\prime) = B\cos\left(\omega^\prime t^\prime + \eta\right)\cos\left(\frac{\omega^\prime}{c}x^\prime + \xi\right)\label{newone}
\end{equation} defined with the variables $t'$ and $x'$. From the analysis made before  we can use this field  to match the motion of a tachyonic particle with velocity  $w_{\rm p}=\frac{c^2}{v_{\rm p}}$. For this we use  in Eq.~\ref{newone} a wave with the same pulsation $\omega'$ as in the badryonic case. However, as we will see  below it implies that we use a different spring with pulsation $\Omega_{\rm p}\neq\omega_{\rm p}$.  Hence, using Eq.~\ref{eq:tachLorTransf} we get in the laboratory frame:
\begin{eqnarray}
u(t,x) = B\cos{\left[-\frac{\omega'}{c}\Gamma_{\rm p}(x-w_{\rm p}t)+\eta\right]}\nonumber\\ \times \cos{\left[-\omega'\Gamma_{\rm p}\left(t-\frac{w_{\rm p}}{c^2}x\right)+\xi\right]}.\label{fieldtac}
\end{eqnarray}
We find that, once again, the roles of two quantities have been swapped.
We have now 
\begin{eqnarray}
\Phi_{\textrm{tach.}}=\omega't'+\eta=-\frac{\omega'}{c}\Gamma_{\rm p}(x-w_{\rm p}t)+\eta\nonumber\\
S_{\textrm{tach.}}=\frac{\omega'}{c}x'+\xi=-\omega'\Gamma_{\rm p}\left(t-\frac{w_{\rm p}}{c^2}x\right)+\xi\label{eq:phasesTach}
\end{eqnarray}
 Here, the carrying phase wave and the envelope have been swapped: the supersonic phase wave that we had in the bradyonic case is now traveling alongside the particle; the group wave is now slower than the particle and has become the phase wave (i.e., $S_{\textrm{tach.}}=\Phi_{\textrm{brad.}}$ and $S_{\textrm{brad.}}=\Phi_{\textrm{tach.}}$). The field itself remains unchanged, and only the roles of its two components with respect to the particle have been changed. Thus, all the results that were obtained in the case of a particle with velocity $v_{\rm p}$ can be used for a tachyonic particle of velocity $w_{\rm p} = c^2/v_{\rm p}$ if we keep in mind the symmetrical nature of this supersonic regime. More precisely, considering a uniform motion $x_{\rm p}(t)=w_{\rm p}t+X_{\rm p,i}$ with $X_{\rm p,i}$ a constant and using the holonomic condition Eq.~\ref{eq:holo} together with the oscillatory $z_{\rm p}(t)$ motion of Eq.~\ref{eq:zHarm} we get by identification with Eq.~\ref{fieldtac}:   
\begin{eqnarray}
\varphi = -\xi - \frac{\omega'\Gamma_{\rm p} X_{\rm p,i}w_{\rm p}}{c^2}+ 2\pi p & (\textrm{with $p\in \mathbb{Z}$}),\nonumber\\
A=B\cos{\left(\frac{\omega'}{c}\Gamma_{\rm p}X_{\rm p,i}-\eta\right)},\nonumber\\
\Omega_{\rm p}=\frac{\omega'}{\Gamma_{\rm p}}=\frac{\omega'\frac{c}{v_{\rm p}}}{\gamma_{\rm p}}.\label{eq:tac}
\end{eqnarray} 
Comparing the frequency-locking condition with the result obtained in Eq.~\ref{eq:clocks} we deduce the constraint \begin{equation}\Omega_{\rm p}\frac{c}{v_{\rm p}}=\omega_{\rm p}\label{tac}\end{equation} and consequently $\Omega_{\rm p}<\omega_{\rm p}$. 

There are, however, a few noteworthy differences in some of the equations, mainly the dispersion relation between the wave pulsation $\Omega=\Omega'\Gamma_{\rm p}$ and wave vector $K=\Omega\frac{w_{\rm p}}{c^2}$ which becomes
\begin{equation}
\frac{\Omega^2}{c^2} - K^2 = - \frac{\Omega^{\prime 2}}{c^2}
\end{equation}
and the Klein-Gordon equation for the wave $\Psi = e^{iS_{\textrm{tach.}}}$ (with $S_{\textrm{tach.}}=Kx-\Omega t+\xi$)
\begin{equation}
\square\Psi = +\frac{\Omega^{\prime 2}}{c^2}\Psi \label{aKG}
\end{equation}
where the signs in front  of $\frac{\Omega^{\prime 2}}{c^2}$ have been reversed compared with $\frac{\omega^{\prime 2}}{c^2}$ in Eq.~\ref{eq:disp} and Eq.~\ref{eq:KG}.  This is specific of tachyonic motions where the pulsation or mass can envisioned as purely imaginary $\tilde{\Omega}=i\Omega'$ (for instance we have $\Omega=\frac{\tilde{\Omega}}{\sqrt{1-\frac{w_{\rm p}^2}{c^2}}}$~\cite{Terletsky1,Bilaniuk,Terletsky2}). Eq.~\ref{aKG} thus equivalently reads $\square\Psi = -\frac{\tilde{\Omega}^{2}}{c^2}\Psi$ which is the usual form for the Klein-Gordon equation  but now with a purely imaginary mass.\\ 
\indent Once more, we point out that our supersonic particle is only superficially looking as a tachyon. Indeed, genuine relativistic tachyons would induce reluctant causality and thermodynamic problems~\cite{Terletsky1,Bilaniuk,Terletsky2,Bell} which are often considered as fatal objections to their mere existence. Here the analogy with tachyon is not complete. 
For instance observe that we can in analogy with the subsonic regime define   the Hamiltonian and linear momentum as $H_{\rm p}:=m_{\rm p}c^2\Gamma_{\rm p}=Q\Omega$ and $P_{\rm p}:=m_{\rm p}w_{\rm p}\Gamma_{\rm p}=QK$.  Physically this corresponds to a tachyonic particle of imaginary mass $im_{\rm p}=iQ\Omega'/c^2=iQ\omega'/c^2$ while the physical `Newtonian' mass is of course $m_{\rm p}$. These expressions are in general clearly different from the usual  kinetic energy  and momentum  of a Newtonian particle.\\
\indent To illustrate the tachyon dynamics we show in  Fig.~\ref{fig:evo2}  an example for this transparency regime (the normalized parameters are indicated in the caption).  Here we have for  the phase wave the temporal period $T_{\textrm{phase}}=\frac{2\pi}{\Omega}=\frac{2\pi}{\omega}\frac{c}{v_{\rm p}}=1$ and a spatial period $\lambda_{\textrm{phase}}=\frac{2\pi}{\Omega}\frac{c^2}{w_{\rm p}}=\frac{2\pi}{\omega}c=0.1$ which have to be compared with the temporal and spatial period of the group wave  $T_{\textrm{group}}=\frac{2\pi}{\Omega}\frac{c}{w_{\rm p}}=\frac{2\pi}{\omega}=0.1\leq T_{\textrm{phase}}$, $\lambda_{\textrm{group}}=\frac{2\pi}{\Omega}c=\frac{2\pi}{\omega}\frac{c}{v_{\rm p}}=1\geq \lambda_{\textrm{phase}}$ (as illustrated in Fig.~\ref{fig:evo2} for $\lambda_{\textrm{phase}}/\lambda_{\textrm{group}}=0.1$. Clearly the role of phase and group waves have been swapped compared to the bradyonic case.  This stresses the strong symmetry existing between our bradyonic and tachyonic particle.\\
\indent Remarkably, it means that a same $u-$wave can carry  several bradyons  of velocity $v_{\rm p}$ associated with a spring of pulsation $\omega_{\rm p}$ and tachyons  of velocity $w_{\rm p}=\frac{c^2}{v_{\rm p}}$ but with a spring of pulsation $\Omega_{\rm p}$ (compare Figs.~\ref{fig:evo1} and \ref{fig:evo2}). The different particles could actually move together on the same string since the transparency condition $N(t)=0$ ensures that the particles do not interact with the wave and completely ignore each other.  Therefore, we have here the possibility to generate collective excitations actually reminiscent of bosons surfing on a given wave. Consider for instance the case of  $n$ particles (bradyons or tachyons) coherently driven by a carrying $u-$wave.   The situation shows some similarity with a Fock state $|n\rangle$ in quantum mechanics for a collective excitation  (e.g., photons or phonons) with $n$ energy quanta.  We could imagine a statistical ensemble of such strings carrying   a various number of quanta  $n=0,1,...,+\infty$  and subsequently develop a  particle bosonic statistics with partition function $Z=\sum_{n=0}^{n=+\infty}e^{-\frac{n \varepsilon_{\omega'}}{K_B T}}$ ($K_B$ is the Boltzmann constant and $T$ the temperature of the ensemble). The energy $\varepsilon_{\omega'}$ is the Hamiltonian $H_{\rm p}$ associated with the bradyons or tachyons characterized by the frequency $\omega'$.  We naturally obtain  Planck's law or more generally and realistically a bosonic statistics if the number of particle is finite and fixed.  Clearly, this issue is very interesting for developing mechanical analogs of quantum statistics and  will deserve further studies in connection with equilibrium and non-equilibrium thermodynamics.    
            
\section{Conclusions}
\label{sec4}
\indent We have developed a simple non-relativistic mechanical analog of de Broglie's wave-particle duality. In the model considered, an oscillating bead or particle mimics an internal quantum clock in phase with a transverse acoustic wave field $u(t,x)$. Despite its non relativistic nature the model is based on a Lorentz transformation where the sound celerity $c$ of waves on the string replaces the light velocity. Therefore the model offers strong similarities with de Broglie's approach which was based on special relativity.  De Broglie named this phase-locking feature the `phase-harmony' \cite{debroglie1927} since it emphasizes the fundamental role of the quantum relation  $m_{\rm p}c^2=\hbar\omega_{\rm p}$. Like with de Broglie's theory the acoustic $u-$field here generates a phase-wave $\Psi$ acting as a guiding field for the particle in pure analogy with  Bohmian interpretation of QM. In other words, in the presented model the $u$-field unifies the particle and the wave in a inseparable structure as summarized by the holonomic condition $z_{\rm p}=u(t,x_{\rm p})$. Thus, one can speak about `wave monism' and in that limited sense our analogy deciphers the meaning of wave-particle duality. Therefore, our work enables to figure out and visualize 
clearly a possible mechanical interpretation of the wave particle
duality and especially of the phase-harmony. It gives a deeper understanding of the possible 
cause for the clock synchronization on which leans de Broglie’s theory.\\
\indent Interestingly our approach which was developed  for both the `bradyonic' (i.~e., subsonic) and `tachyonic' (i.~e., supersonic) regime is \textit{a priori} not limited to the uniform motions. Indeed, the set of equations~\ref{total1}, \ref{total2}, and \ref{total3} can easily be extended in order to include the effects of a more complex potential $V(t,x_{\rm p},z_{\rm p})$. For example,  by adding a longitudinal potential  $V(t,x_{\rm p})$ Eq.~\ref{total1} becomes 
\begin{eqnarray}
m_{\rm p}\ddot{x}_{\rm p}(t)=-\partial_{x} u(t,x)|_{x=x_{\rm p}(t)}\, N(t)-\partial_xV(t,x)|_{x=x_{\rm p}(t)}\nonumber\\
\end{eqnarray}
 whereas  Eqs.~\ref{total2}, and \ref{total3} are let unaffected. The model could thus in principle be applied to regimes involving confining potentials which are of particular interest for further studies on mechanical analogs of QM.  
 In that context, the development of the numerical method employed for generating the movie (see \cite{Movie}) will bring further insight into the complex dynamic  of the system (especially when the `transparency' condition $N(t)=0$ is no longer valid). 
 Finally,  while the basis of our mechanical model is particularly simple, the complexity of the various dynamics  (with $N=0$ and $N\neq 0$) open interesting and fundamental questions.
We think in particular of the unexplored tachyonic regime as well as the highly likely chaotic regimes that can also
be addressed experimentally.
For all these reasons, we hope  that this work will stimulate  further studies exploiting the potential of de Broglie's mechanical interpretation of quantum physics.
\acknowledgments
The authors wish to thank J. Duplat for helpful discussions and comments. Authors also acknowledge the continuous support of Neel institute during this project.
C.~Poulain thanks the CEA-Leti for its support.


\begin{thebibliography}{99}
\bibitem{debroglie1925}
L. de Broglie,  Ann. Phys. \textbf{10}, 22-128 (1925).
\bibitem{debroglie1926}
L. de Broglie, Nature \textbf{118}, 441-442 (1926).
\bibitem{debroglie1927}
L. de Broglie, J. Phys. Radium \textbf{8} 225-241 (1927)  translated in: de L. Broglie and L. Brillouin  \textit{Selected papers on wave mechanics} (Blackie and Son, Glasgow, 1928). 
\bibitem{debroglie1987}
L. de Broglie, Ann. Fond. de Broglie \textbf{12} 1-23 (1987).
\bibitem{Madelung1927}
E. Madelung, Z. Phys. \textbf{40}, 322–326 (1927).
\bibitem{Valentini}
G. Bacciagaluppi A. Valentini, {\it Quantum theory at the crossroads: Reconsidering the 1927 Solvay Conference} (Cambridge Univ. Press, Cambridge, 2009). 
\bibitem{Bohm1952}
D. Bohm,Phys. Rev. \textbf{85}, 166–179 (1952).
\bibitem{Hiley}
D. Bohm and B.~J.~Hiley,  \textit{The undivided Universe} (Routledge, London, 1993). 
\bibitem{Couder2005}
Y. Couder S. Proti\`ere, E. Fort, Nature \textbf{437}, 208 (2005).
\bibitem{Couder2006}
Y. Couder and E. Fort, Phys. Rev. Lett. \textbf{97}, 154101 (2006).
\bibitem{Bush2013}
D. M. Harris, J. Moukhtar, E. Fort, Y. couder and J.~W.~M Bush, Phys. Rev. E \textbf{88}, 011001(R). 
\bibitem{Bush2015}
J.~W.~M.~Bush J, Phys. Today \textbf{68} 47-53 (2015).
\bibitem{Rayleigh}
Lord Rayleigh, \textit{The theory of sounds} (University of Michigan library, 1894) reprinted by Dover publications 1945. 
\bibitem{Helmholtz}
H.~V.~Helmholtz,\textit{Vorlesungen \"uber theoretische  physik, Vol 3} (Ambrosuis Barth, Leipzig, 1908).  
\bibitem{Morse}
P.~M.~Morse and K.~U.~Ingard, \textit{Theoretical Acoustics} (Princeton University Press, Princeton, 1992).
\bibitem{Boudaoud1999}
A. Boudaoud, Y. Couder and M. Ben Amar, Eur. J. Phys. B \textbf{9}, 159-165 (1999).
\bibitem{Borghesi2017}
C. Borghesi, Ann. Fond. de Broglie \textbf{42}, 161-196 (2017).
\bibitem{debroglie1925b}
L. de Broglie,  C. R. Acad. Sci. \textbf{180}, 498-500 (1925).
\bibitem{Duplat}
P.~Y.~Gires, J.~Duplat, A.~Drezet, C.~Poulain, EPL \textbf{127}, 34002 (2019).
\bibitem{Poincare}
H. Poincar\'e, \textit{La th\'eorie de Lorentz et le principe de r\'eaction in Recueil des travaux offerts par les auteurs \`a H. A. Lorentz}, pp. 252-278 (Nijhoff, The Hague, 1900).  
\bibitem{voigt}
W. Voigt Voigt, G\"ottinger Nachrichten \textbf{8}, 177–238 (1887).
\bibitem{debroglie1925c}
L. de Broglie, \textit{Ondes et mouvements} (Gauthier Villars, Paris 1926).
\bibitem{Movie} Follow this link to see the attached movie XXXXX.
\bibitem{Loustau}
J. Loustau, \textit{Numerical differential equations} (World scientific press, Singapore, 2016).
\bibitem{Terletsky1}
J.~P. Terletsky, J. Phys. Radium \textbf{21}, 681-684 (1960).
\bibitem{Bilaniuk}
O.~M.~P.~Bilaniuk, V.~K.~Deshpande and E.~C.~G. Sudarshan, Am. J. Phys. \textbf{30},  718-723 (1962).
\bibitem{Feinberg}
G.~Feinberg, Phys.~Rev. 159, 1089-1105 (1967). 
\bibitem{Terletsky2}
J.~P. Terletsky, J. Phys. Radium \textbf{23}, 910-920 (1962).
\bibitem{Bell}
J.~S.~Bell, \textit{`La nouvelle cuisine' in Between science and technology, North-Holland Delta Series , edited by A. Sarlemijn and P.Kroes}, pp. 97-115 (Elsevier, Amsterdam, 1990).
\end{thebibliography}
\end{document}